\newif\ifJOURNAL
\JOURNALfalse
\newif\ifWP
\WPfalse
\newif\ifBASIC
\BASICfalse

\newif\ifFULL
\FULLfalse
\newif\ifLATIN
\LATINfalse

\BASICtrue

\LATINtrue	

\newif\ifnotJOURNAL		
\notJOURNALtrue
\ifJOURNAL\notJOURNALfalse\fi

\newif\ifnotWP		
\notWPtrue
\ifWP\notWPfalse\fi

\newif\ifnotFULL	
\notFULLtrue
\ifFULL\notFULLfalse\fi

\newif\ifnotLATIN	
\notLATINtrue
\ifLATIN\notLATINfalse\fi

\ifnotLATIN
  
\fi
\ifLATIN
  
\fi

\ifJOURNAL
  \documentclass{article}
  \usepackage{amsmath,amsfonts,amssymb,amsthm,latexsym,graphicx,epigraph}
  \newcommand{\Extra}[1]{}
\fi

\ifWP
  \documentclass[12pt]{article}
  \usepackage{amsmath,amsfonts,amssymb,amsthm,latexsym,graphicx,epigraph}
  \input{/Doc/Computing/Latex/gt.txt}
  \newcommand{\Extra}[1]{}
\fi

\ifBASIC
  \documentclass[10pt]{article}
  \usepackage{amsmath,amsfonts,amssymb,amsthm,latexsym,graphicx,epigraph}
  \newcommand{\Extra}[1]{}
\fi

\ifFULL
  \usepackage{color}
  \renewcommand{\Extra}[1]{\blue{#1}}
  
  \newcommand{\blue}[1]{\textcolor{blue}{#1}}
  \newcommand{\bluebegin}{\begingroup\color{blue}}
  \newcommand{\blueend}{\endgroup}

\fi

\allowdisplaybreaks[4]

\newcommand{\Vladimir}{Vladimir}

\ifnotLATIN
  \input{OT2enc.def}
  
  \usepackage{CJK}
\fi

\newcommand{\dd}{\mathrm{d}}		
\newcommand{\tr}{^{\textrm{\tiny T}}} 

\newcommand{\K}{\mathcal{K}}		

\newcommand{\FFF}{\mathcal{F}}		

\newcommand{\eqdef}{\mathrel{:=}}     

\newcommand{\bbbr}{\mathbb{R}}			

\newcommand{\bbbp}{\mathbb{P}}			

\DeclareMathOperator{\ProbPh}{\bbbp}    
\DeclareMathOperator{\ProbRN}{\tilde\bbbp}  

\newcommand{\bbbe}{\mathbb{E}}			

\DeclareMathOperator{\ExpectPh}{\bbbe}    

\newtheorem{theorem}{Theorem}

\theoremstyle{definition}

\newtheorem{remark}{Remark}

\title{A simplified Capital Asset Pricing Model}

\ifJOURNAL
  \author{Vladimir Vovk}
\fi

\ifWP
  \author{Vladimir Vovk}
  
\fi

\ifBASIC
  \author{Vladimir Vovk}
\fi

\begin{document}
\maketitle

\begin{abstract}
  We consider a Black--Scholes market
  in which a number of stocks and an index are traded.
  The simplified Capital Asset Pricing Model is the conjunction
  of the usual Capital Asset Pricing Model, or CAPM,
  and the statement that the appreciation rate of the index
  is equal to its squared volatility plus the interest rate.
  (The mathematical statement of the conjunction is simpler than that of the usual CAPM.)
  Our main result is that either we can outperform the index
  or the simplified CAPM holds.
\end{abstract}

\epigraph{Simply buying and holding the stocks in a broad-market index
  is a strategy that is very hard for the professional portfolio manager
  to beat.}
{Burton G. Malkiel \cite{malkiel:1999}}

\section{Introduction}
\label{sec:introduction}

The simplified CAPM (SCAPM)
says that the market price of risk coincides (at least approximately)
with the volatility of the index.
This note gives two formalizations of the following disjunction:
either SCAPM holds or we can outperform the index.
The formalizations are quantitative,
in that they characterize the tradeoff
between the degree to which we can outperform the market
and the discrepancy between the market price of risk and the volatility of the index.
One formalization (Theorem~\ref{thm:asymptotic})
says, in the case of a constant-coefficient market,
that asymptotically we can almost surely outperform the index by 50\% per period
times the squared norm of the discrepancy.
The other formalization (stated in Section~\ref{sec:proof}) is non-asymptotic:
it says that for a finite investment horizon $T$
we can beat the index by a large factor
with a probability close to 1
unless the discrepancy has the order of magnitude $T^{-1/2}$
(with a given constant in front of $T^{-1/2}$).

There are two natural interpretations of our results.
If we believe that the index is efficient,
in that we do not expect a prespecified (and very simple) trading strategy
to outperform the index (cf.\ the epigraph above),
we can conclude that the SCAPM holds.
And if we do not believe that the SCAPM holds,
we can outperform the index.

Another statement of the SCAPM is that the appreciation rate of a security
exceeds the interest rate by the covariance
between the volatilities of the security and the index.
This implies not only a version of the standard CAPM
but also the appreciation rate of the index
being the sum of its squared volatility and the interest rate.

Our main result is mathematically very simple, almost trivial:
its proof is little more that an application of the identity
$a^2+b^2-2ab=(a-b)^2$.

\section{Main result and its discussion}

Consider a financial market in which $K+1$ securities, $K\ge0$, are traded:
an index and $K$ stocks.
The time interval is $[0,\infty)$;
we will be using the framework of \cite{karatzas/shreve:1998}, Section~1.7.
The price of the index at time $t$ is $S^0_t$,
and the price of the $k$th stock, $k=1,\ldots,K$, is $S^k_t$.
Suppose the prices satisfy the multi-dimensional Black--Scholes model
\begin{equation}\label{eq:physical}
  \frac{\dd S^k_t}{S^k_t}
  =
  \mu^k_t\dd t + \sigma_t^{k,1}\dd W^1_t + \cdots + \sigma_t^{k,D}\dd W^D_t,
  \quad
  k=0,\ldots,K,
\end{equation}
where $(W^1_t,\ldots,W^D_t)=W_t$ is a standard Brownian motion in $\bbbr^D$.
Set $\mu_t\eqdef(\mu^0_t,\ldots,\mu^K_t)\tr$
(this is the \emph{appreciation vector})
and $\sigma_t^k\eqdef(\sigma^{k,1}_t,\ldots,\sigma^{k,D}_t)\tr$ for $k=0,\ldots,K$
(these are the \emph{volatility vectors},
also called \emph{volatilities} in Section~\ref{sec:introduction}),
and let $\sigma_t$ be the $(K+1)\times D$ matrix $\sigma^{k,d}_t$,
$k=0,\ldots,K$, $d=1,\ldots,D$.
The interest rate at time $t$ is denoted $r_t$.

We make the assumptions of \cite{karatzas/shreve:1998}, Section~1.7,
except that we allow $D<K+1$.
We assume that $D\le K+1$ and that $\sigma_t$ is a full-rank matrix
for almost all $t$
(we always consider the Lebesgue measure on $t$)
almost surely,
which makes the market complete
(\cite{karatzas/shreve:1998}, Theorem~1.6.6 and Remark~1.4.10).
Assuming that our market is viable,
we obtain the existence of a \emph{market price of risk} process $\theta_t$ in $\bbbr^D$
such that for almost all $t$ we have
$\mu_t-r_t\boldsymbol{1}=\sigma_t\theta_t$ a.s.
In the interpretation of our results,
we will usually assume that $D=K+1$,
in which case $\theta_t=\sigma_t^{-1}(\mu_t-r_t\boldsymbol{1})$
for almost all $t$ almost surely.
Another interesting case is where $D=K$:
this arises naturally when our $K$ stocks are all the stocks traded in the market
and the index is their capital-weighted average.

Set $R_t\eqdef\exp(\int_0^t r_s\dd s)$;
in particular, $R_0=1$.
For simplicity and without loss of generality we assume
$S^k_0=1$, for all $k$.
We consider only admissible trading strategies
(as defined in \cite{musiela/rutkowski:2005}, Definition~3.1.4).
If $A$ and $B$ are two events,
$A\Longrightarrow B$ stands for $A^c\cup B$.

\begin{theorem}\label{thm:asymptotic}
  There exists a nonnegative wealth process $\K_t$ such that $\K_0=1$ and
  \begin{equation}\label{eq:asymptotic-1}
    \int_0^{\infty}\left\|\theta_t-\sigma_t^0\right\|^2\dd t
    =
    \infty
    \Longrightarrow
    \lim_{t\to\infty}
    \frac
      {\ln\K_t - \ln S^0_t}
      {\int_0^t\left\|\theta_s-\sigma_s^0\right\|^2\dd s}
    =
    \frac12
  \end{equation}
  almost surely.
\end{theorem}

In the discussion in the rest of this section
we will consider the constant-coefficient market,
in which $r_t$, $\mu^k_t$, and $\sigma^{k,d}_t$ do not depend on $t$,
and so we will drop the subscript $t$.
If we believe that the index is ``asymptotically efficient'',
in that we do not expect to be able to outperform it even in the sense of
$$
  \limsup_{t\to\infty}
  \frac{\ln\K_t - \ln S^0_t}{t}
  >
  0,
$$
we have $\theta=\sigma^0$,
i.e., $\mu-r\boldsymbol{1}=\sigma\sigma^0$.
In other words, we have
\begin{equation}\label{eq:SCAPM}
  \mu^k = r + \sigma^k \cdot \sigma^0,
  \quad
  k=0,\ldots,K.
\end{equation}
We call the set of equalities (\ref{eq:SCAPM})
the \emph{simplified CAPM} as they are the conjunction of the version
\begin{equation*} 
  \mu^k = r + \frac{\sigma^k \cdot \sigma^0}{\left\|\sigma^0\right\|^2}(\mu^0-r),
  \quad
  k=0,\ldots,K,
\end{equation*}
of the standard CAPM (see, e.g., \cite{fama/french:2004}, pp.~28--29)
and the expression
\begin{equation*} 
  \mu^0 - r = \left\|\sigma^0\right\|^2
\end{equation*}
for the equity premium $\mu^0-r$.

\begin{remark}\label{rem:unnatural}
  Constant-coefficient markets are mathematically consistent
  and convenient for illustrating the meaning of our results,
  but they are somewhat unnatural:
  if all stocks in the market have constant appreciation rates and volatilities,
  the capital-weighted average of all stocks
  may not have constant appreciation rate and volatility;
  and many well-known indexes are defined as capital-weighted averages of stocks.
\end{remark}

A weakness of Theorem~\ref{thm:asymptotic} is its asymptotic character;
however, its proof in the next section
will show that there is nothing asymptotic
in the phenomenon that Theorem~\ref{thm:asymptotic} expresses.

\section{Proof of Theorem~\ref{thm:asymptotic}}
\label{sec:proof}

By Girsanov's theorem,
\begin{equation}\label{eq:tilde-physical}
  \tilde W_t
  \eqdef
  W_t
  +
  \int_0^t \theta_s \dd s
\end{equation}
is a standard Brownian motion under a new probability measure $\ProbRN$
called the \emph{risk-neutral measure} \cite{karatzas/shreve:1998}.
Our notation for the physical measure (\ref{eq:physical}) will be $\ProbPh$;
the restrictions of $\ProbPh$ and $\ProbRN$ to $\FFF^{(T)}(t)$
(in the notation of \cite{karatzas/shreve:1998})
will be denoted $\ProbPh^T_t$ and $\ProbRN^T_t$, respectively.
Girsanov's theorem also gives
$$
  \frac{\dd\ProbPh^T_t}{\dd\ProbRN^T_t}
  =
  \exp
  \left(
    \int_0^t
    \theta_s
    \cdot
    \dd\tilde W_s
    -
    \frac12
    \int_0^t
    \left\|
      \theta_s
    \right\|^2
    \dd s
  \right).
$$
For the nonnegative wealth process
$\K_t\eqdef R_t\frac{\dd\ProbPh^T_t}{\dd\ProbRN^T_t}$
(which, as we can see, does not depend on $T$)
we have
\begin{align*}
  \ln\K_t
  &=
  \int_0^t r_s \dd s
  +
  \int_0^t
  \theta_s
  \cdot
  \dd\tilde W_s
  -
  \frac12
  \int_0^t
  \left\|
    \theta_s
  \right\|^2
  \dd s\\
  &=
  \int_0^t r_s \dd s
  +
  \int_0^t
  \theta_s
  \cdot
  \dd W_s
  +
  \frac12
  \int_0^t
  \left\|
    \theta_s
  \right\|^2
  \dd s,
\end{align*}
which in conjunction with the strong solution
\begin{equation*} 
  S^k_t
  =
  \exp
  \left(
    \int_0^t \mu^k_s \dd s
    -
    \frac12
    \int_0^t
    \left\|\sigma^k_s\right\|^2
    \dd s
    +
    \int_0^t\sigma^k_s \cdot \dd W_s
  \right),
  \quad
  k=0,\ldots,K,
\end{equation*}
to (\ref{eq:physical}) (for $k=0$) gives
\begin{multline}\label{eq:central}
  \ln\K_t-\ln S^0_t
  =
  \int_0^t
  (r_s-\mu^0_s)
  \dd s
  +
  \frac12
  \int_0^t
  \left\|
    \theta_s
  \right\|^2
  \dd s
  +
  \frac12
  \int_0^t
  \left\|
    \sigma^0_s
  \right\|^2
  \dd s\\
  +
  \int_0^t
  (\theta_s-\sigma^0_s)
  \cdot
  \dd W_s
  =
  \frac12
  \int_0^t
  \left\|
    \theta_s - \sigma^0_s
  \right\|^2
  \dd s
  +
  \int_0^t
  (\theta_s-\sigma^0_s)
  \cdot
  \dd W_s
\end{multline}
(the second equality using $r_s-\mu^0_s=-\sigma^0_s\cdot\theta_s$).
By the law of the iterated logarithm and the Dubins--Schwarz theorem,
we can see that (\ref{eq:asymptotic-1}) can in fact be strengthened to
\begin{multline*}
  \int_0^{\infty}\left\|\theta_t-\sigma_t^0\right\|^2\dd t
  =
  \infty
  \Longrightarrow\\
  \lim_{t\to\infty}
  \frac
    {\ln\K_t - \ln S^0_t - \frac12 \int_0^t\left\|\theta_s-\sigma_s^0\right\|^2\dd s}
    {
      \sqrt
      {
        2
        \int_0^t\left\|\theta_s-\sigma_s^0\right\|^2\dd s
	\ln\ln\int_0^t\left\|\theta_s-\sigma_s^0\right\|^2\dd s
      }
    }
  =
  1
  \quad
  \text{a.s.}
\end{multline*}

\section{A finite-horizon implication}
\label{sec:exact}

Let us see what (\ref{eq:central}) gives
in the case of a constant-coefficient market and a finite investment horizon $T>0$.
Set $D:=\theta-\sigma^0$
(this is the discrepancy that we discussed in Section~\ref{sec:introduction}).
Fix constants $\epsilon,\delta\in(0,1)$ (the interesting case is where they are small).
Since
$$
  \ln\K_T - \ln S^0_T
  =
  \frac12 \left\|D\right\|^2T + D\cdot W_T,
$$
the probability that
$$
  \ln\K_T - \ln S^0_T
  \le
  \ln\frac{1}{\delta}
$$
is less than $\epsilon$
if and only if
$$
  \frac12 \left\|D\right\| \sqrt{T}
  -
  \frac{1}{\left\|D\right\|\sqrt{T}}
  \ln\frac{1}{\delta}
  \ge
  z_{\epsilon},
$$
where $z_{\epsilon}$ stands
for the upper $\epsilon$-quantile of the standard normal distribution.
Solving this quadratic inequality,
we can see that $\K_t/S_t^0>1/\delta$
with probability at least $1-\epsilon$ unless
\begin{equation}\label{eq:weak}
  \left\|D\right\|
  <
  \frac{z_{\epsilon}+\sqrt{z^2_{\epsilon}+2\ln\frac{1}{\delta}}}{\sqrt{T}}
  <
  \frac{2z_{\epsilon}+\sqrt{2\ln\frac{1}{\delta}}}{\sqrt{T}}.
\end{equation}
In other words,
our strategy beats the index by a factor of more than $1/\delta$
with probability at least $1-\epsilon$
unless the approximate SCAPM (\ref{eq:weak}) holds.

Replacing the wealth process $\K_t$ by an \emph{ad hoc} wealth process
(depending on $\epsilon$ and $\delta$),
it is possible to improve (\ref{eq:weak}) to
\begin{equation*}
  \left\|D\right\|
  \le
  \frac{z_{\epsilon}+z_{\delta}}{\sqrt{T}}
\end{equation*}
(see \cite{vovk:arXiv1109CAPM}, Theorem~9.2).

\section{Connection with the optimal growth rate of wealth}

In the case of the model (\ref{eq:physical}) with constant coefficients
(including the interest rate),
the SCAPM can be easily deduced from the known results
about the optimal growth rate of wealth.
According to Corollary~3.10.2 in \cite{karatzas/shreve:1998},
the optimal growth rate $\limsup_{t\to\infty}\frac{1}{t}\ln\K_t$
is $r+\frac12\left\|\theta\right\|^2$ almost surely.
Since security $k$ (including the index) cannot grow faster than the optimal portfolio,
$$
  \mu_k - \frac12\left\|\sigma_k\right\|^2
  \le
  r + \frac12\left\|\theta\right\|^2.
$$
The difference between the two sides of this inequality is
$$
  r + \frac12\left\|\theta\right\|^2
  -
  \mu_k + \frac12\left\|\sigma_k\right\|^2
  =
  \frac12\left\|\theta\right\|^2 + \frac12\left\|\sigma_k\right\|^2
  -
  \sigma_k \cdot \theta
  =
  \frac12
  \left\|\theta-\sigma_k\right\|^2.
$$
For an asymptotically efficient index, we have $\theta=\sigma_0$.
The shortfall of the growth rate of stock $k$ is
$
  \frac12
  \left\|\theta-\sigma_k\right\|^2
$;
this was called the \emph{theoretical performance deficit}
in \cite{vovk/shafer:2008CAPM} and \cite{vovk:arXiv1109CAPM}
as the shortfall can be attributed to insufficient diversification
as compared to the index.

\subsection*{Acknowledgements}

Thanks to Glenn Shafer, Robert Merton, Wouter Koolen, and Jan Ob\l\'oj
for their help and advice.


\end{document}